\documentclass[conference]{IEEEtran}
\IEEEoverridecommandlockouts
\usepackage{cite}
\usepackage{url}
\usepackage{amsmath,amssymb,amsfonts}
\usepackage[ruled]{algorithm2e}
\usepackage{graphicx}
\usepackage{textcomp}
\usepackage{xcolor}
\usepackage{chngcntr}
\counterwithout{table}{section}
\def\BibTeX{{\rm B\kern-.05em{\sc i\kern-.025em b}\kern-.08em
    T\kern-.1667em\lower.7ex\hbox{E}\kern-.125emX}}

\IEEEoverridecommandlockouts

\usepackage{cite}
\usepackage{amsmath,amssymb,amsfonts}
\usepackage{algorithmic}
\usepackage{graphicx}
\usepackage{textcomp}
\usepackage{xcolor}

\begin{document}

\title{PSketch: A Priority-Aware Sketch Architecture
for Real-Time Flow Monitoring via eBPF \\
\vspace{-1.5em}
{\footnotesize \textsuperscript{*}}}    
\author{
    Yuanjun Dai\textsuperscript{1}, Qingzhe Guo\textsuperscript{1}, Xiangren Wang\textsuperscript{2} \\
    \textsuperscript{1} Case Western Reserve University, Cleveland, OH, USA \\
    \textsuperscript{2} University of Florida, Gainesville, FL, USA \\
    Contact: yxd429@case.edu
}

\vspace{-12em}
\maketitle
\begin{abstract}
Sketch-based monitoring in SDN often suffers from tightly coupled pipeline and memory constraints, limiting algorithmic flexibility and reducing accuracy. We propose PSketch, the first in-kernel priority-aware sketching framework implemented with eBPF. It ensures lossless tracking of high-priority flows via a hash-based table and approximates top-$k$ elephant flows using a sketch pipe. PSketch supports both TCP and UDP and enables in-kernel retransmission tracking with minimal overhead. Unlike SDN-based approaches, it runs on commodity Linux systems, removing hardware dependencies. We perform evaluation on 10 Gbps CAIDA traces. Result shows that PSketch achieves 96.0\% top-$k$ detection accuracy, and 96.4\% retransmission recall and only 0.7\%  of throughput degration.
\end{abstract}

\vspace{-0.5em}
\section{Introduction}
In modern network management, network flow monitoring is critical for traffic engineering, congestion control, and system optimization~\cite{huang2017sketchvisor, kim2023robust}. This challenge has become particularly critical in the era of large-scale distributed machine learning (DML), where training modern AI models requires intensive communication across hundreds or thousands of GPU/CPU nodes\cite{lin2022aiacc}.

The latest platforms like FlexFlow and HetPipe\cite{jia2019flexflow} based on flow monitoring can dramatically improve throughput and system efficiency. In addition, researchers have observed that during the forward pass of DML training, the network enters a silent period (no packet transmission) because the gradients are not ready to transmit\cite{dai2023dnn}. This behavior suggests opportunities to interleave the communication of other jobs to improve overall network utilization. Such optimizations are based on timely and fine-grained flow-level telemetry, including the monitoring of DML flows and detection of large-volume elephant flows.

Sketch-based telemetry techniques have proven to be highly effective for scalable real-time monitoring of network traffic\cite{yang2018elastic}.
However, existing sketch-based telemetry tools often depend on programmable switches or hardware-defined data planes. These platforms impose several limitations in practice, such as fixed pipeline architecture, limited on-chip memory, and no support for dynamic data structures~\cite{lin2022mc}. Furthermore, many academic platforms\cite{Duplyakin2019TheDI}, commercial cloud environments\cite{googlecloud}, HPC clusters\cite{OhioSupercomputerCenter}, or legacy data centers\cite{Berman2014GENI} do not deploy or make such specialized hardware available to users. As a result, despite their potential, sketch-based monitoring systems remain difficult to adopt widely.

To address these challenges, we present \textbf{PSketch}, a real-time, priority-aware sketch framework implemented entirely using eBPF in the Linux kernel. PSketch combines a high-resolution priority table for selected flows with a scalable, multi-layer sketch pipeline for estimating elephant flow and cardinalities supporting both TCP and UDP. To provide more informative statics for the framework, we extend traditional sketch capabilities by analyzing per-flow TCP sequence numbers to estimate retransmission rates. This enhancement enables more representative insights into the current network conditions. PSketch does not require specialized hardware, making it suitable for both production and experimental environments. We evaluated our system using 10 Gbps CAIDA\cite{caida2016traces} backbone traces and demonstrated its effectiveness as lightweight and powerful flow telemetry solution.
\section{Related Work}
Recent years have witnessed growing interest in leveraging eBPF for in-kernel network telemetry and measurement. Several studies have explored the feasibility of implementing sketch-based data structures in the kernel. Miano et al.~\cite{miano2023fast} provided a comprehensive evaluation of sketch algorithms in eBPF, showing that careful optimization enables high performance despite instruction and helper limitations. Zang et al.~\cite{zangkernel} applied a Count-Min sketch in eBPF for volumetric DDoS detection, focusing on accurate volume estimation for fixed detection targets. Along similar lines, Yu et al.~\cite{yu2022sketchflow} introduced SketchFlow, a system that leverages hybrid sketches for flow-size estimation, demonstrating the trade-offs between accuracy and memory efficiency in high-speed monitoring.

Beyond basic sketching, subsequent research has emphasized efficiency, flexibility, and adaptivity of kernel-level monitoring. Abranches et al.~\cite{abranches2021xdp} proposed eBPF+XDP primitives for efficient in-kernel monitoring, which reduce resource consumption by triggering user-space involvement only when necessary. Magnani et al.~\cite{magnani2022adaptive} designed a runtime control plane enabling fully adaptive monitoring with eBPF, allowing injection and modification of monitoring logic at runtime with negligible overhead. Zhang et al.~\cite{zhang2021ebpfmon} further demonstrated that adaptive, priority-aware monitoring policies can be realized by dynamically reconfiguring eBPF programs, highlighting the feasibility of fine-grained telemetry under diverse workloads.

In parallel, eBPF has been increasingly integrated into cloud-native observability ecosystems. Open-source frameworks such as Skydive and libebpfflow~\cite{opensource2022ebpf} provide network-flow visibility and transport-level metrics directly from the kernel. OpenTelemetry-eBPF projects~\cite{opentelemetry2022} extend these capabilities into standardized observability pipelines, while Kubernetes-oriented platforms such as Pixie~\cite{pixie2022} showcase the practicality of automatic, code-free data collection across application and network layers. Other systems, such as FlowRadar~\cite{liu2016flowradar}, have also illustrated the potential of combining sketch-based telemetry with programmable data planes to capture fine-grained flow statistics at scale.

PSketch builds on this growing body of work by introducing a novel hybrid design that unifies prioritized monitoring of elephant flows with accurate tracking of large-volume traffic across TCP and UDP. In contrast to prior studies, which often focused on fixed-function sketches or single-purpose monitoring tools, PSketch provides a generalized framework for priority-aware and large-volume flow monitoring. Importantly, it remains fully deployable on commodity Linux systems, offering a practical path to extending sketch-based telemetry capabilities in real-world environments.

\section{Methodology}
\begin{figure}[htbp]
    \centering
    \includegraphics[width=0.49\textwidth]{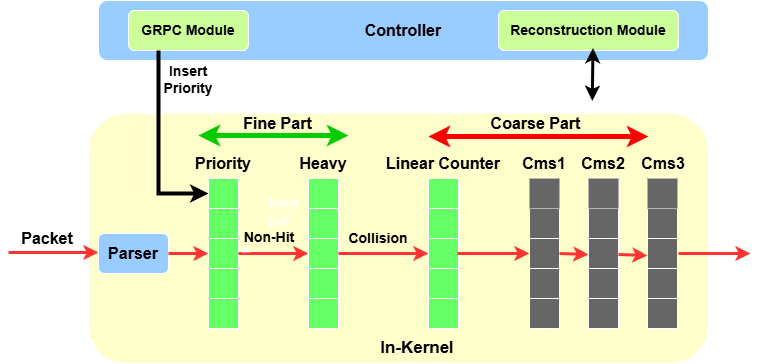}
    \caption{Architecture of PSketch}
    \label{fig:architecture}
\end{figure}
Figure~\ref{fig:architecture} illustrates the overall architecture of our PSketch system, which includes a user-level control component and a kernel-level sketching module.
In order to losslessly monitor selected high priority flows and estimate heavy flow at the same time, PSketch needs to maintain the following data structure. 
    \subsection{Priority Table}
The priority table is implemented using the \texttt{BPF\_HASH} map, which provides \(O(1)\) performance for both insertion and lookup operations. 
In our design, packets should be distinguished as either priority or non-priority flows when they are processed by the priority table. 
To enable this functionality, the 5-tuple of each priority flow must be inserted into the priority table through the controller and then initialize its corresponding entry within the table.
Afterwards, the priority table will check whether the incoming flow has a corresponding entry based on its 5-tuple during packet processing. 
If a match is found, the packet is identified as part of a priority flow and will be precisely monitored.
Otherwise, the packet is classified as a non-priority flow and directed to the subsequent sketch-based structure. 
The existence check enables lightweight, real-time flow classification on the fly.
By using native hash table operations for classification, our method avoids the overhead of explicitly labeling each packet with a priority tag, which is required by SDN-centric approaches\cite{lin2022mc}. Additionally, it lowers the per-packet processing cost and memory requirements, resulting in a more efficient and scalable design.

Although the flow classification method is simple and effective, a major challenge arises in acquiring the complete five-tuple details in advance. Our design addresses this by emphasizing \textit{flexibility}: it provides the core mechanism for priority-aware monitoring while leaving the specific control policy open to customization. Thanks to the programmability of eBPF, a user-level controller can dynamically manage this process, integrating with lightweight communication mechanisms such as gRPC to update monitoring configurations on demand.  

For example, once a target flow (e.g., a DML training flow) is established, its five-tuple information can be immediately transmitted to the controller and inserted into the kernel’s priority table. To ensure the insertion happens before the flow actually occurs, we leverage gRPC’s flexibility to design a confirmation mechanism: the controller only sends an acknowledgment to the application once all keys have been successfully installed, allowing the application to proceed safely. This separation of mechanism and policy enables users to implement their own logic---such as triggering updates based on retransmission rate thresholds, anomaly detection outputs, or application-specific signals---while relying on our system as the foundational in-kernel framework.  

Moreover, because gRPC is lightweight and the five-tuple information is very small, the transmission overhead is negligible. Even under congestion, the confirmation mechanism ensures correctness without introducing noticeable delays. In our experiments, we adopted a fixed policy for simplicity, but in practice, many different policies can be built on top of this design. In the camera-ready version, we will clarify the original rationale of our controller module to make the intended separation between mechanism and policy more explicit.  

Furthermore, a full five-tuple is not strictly necessary---our design allows flexible key selection (e.g., using only the source and destination IP pair) to enable broader traffic visibility when precise identification is not required.

\subsection{Sketch Pipeline}
When a packet is classified as low priority, it will be transmitted to the next sketch pipeline to identify the \textit{Top}-$K$ elephant flows. This pipeline is built around three key components.
\subsubsection{Heavy Flow Table}
The heavy flow table is implemented using a fixed-size \texttt{BPF\_ARRAY}. Each incoming packet is assigned an entry in the array via a hash function. To reduce the possibility of hash collisions, we use the Jenkins hash function\cite{6781990}, which offers a great trade-off between randomness and performance. The input to the hash function is the packet's five-tuple, which attempts to ensure that each flow is mapped to a unique index in the array. 
The table maintains data including the 5-tuple, packet count, next expected sequence number, timestamp, retransmission count, kick-out flag, and negative voting count. It also performs three key operations that collectively support efficient detection and tracking of top-$k$ elephant flows with $\mathcal{O}(N)$ time complexity.
\begin{itemize}
\item \emph{Matching.}
First, we check if the slot is occupied when a packet arrives by looking at whether its \texttt{src\_ip} is \texttt{null}. If the slot is already occupied, we compare the 5-tuple of the packet with the stored flow key. If they match, the packet is considered to be part of the same flow, and its statistics are updated accordingly.

For TCP packets, we compared its sequence number with the next expected sequence number stored in the slot. A smaller current packet's sequence number may indicate a retransmission. To avoid false positives from TCP reordering, we verify if the current packet \texttt{timestamp} exceeds the \texttt{timestamp} in the slot with a threshold (e.g., 3 ms). If both criteria are satisfied, the packet is considered a retransmission and the retransmission counter is incremented. If not, the slot is updated with the latest next expected sequence number and timestamp from the packet for future checks. For UDP flows, the packet count is incremented by 1, and there is no need for a retransmission check.
\item \emph{Collision.}
When a hash collision occurs, we employ a voting-based mechanism to determine whether the flow in the slot should be evicted. Each entry in the heavy flow table maintains a \texttt{negative\_counter}, which records the number of packets that collide with the existing flow's hash slot but do not match its five-tuple.

This counter serves as a simple voting metric: every time a colliding packet arrives, the \texttt{negative\_counter} of the current entry is incremented by 1. 
If the product of \texttt{negative\_counter} and the predefined threshold \texttt{VOTE\_THRE} is less than the flow's \texttt{packet\_count}, we increment the \texttt{negative\_counter} by 1 and forward the colliding packet to the sketch pipeline without eviction.
\item \emph{Kick Flag.}
If the product of \texttt{negative\_counter} and a predefined threshold \texttt{VOTE\_THRE} exceeds the flow’s \texttt{packet\_count}, we consider the flow in the slot outdated. The new flow then evicts the old one and takes over the slot.

To record this event, we set a \texttt{kick\_flag} to \texttt{true}, indicating that the slot has been reassigned due to voting and partial packets of the current flow have been forwarded to a downstream sketch structure. \texttt{Kick\_flag} facilitates later analysis as it signals that the current entry contains only a fragment of the lifetime statistics of the flow.
\end{itemize}
\subsection{Linear Counter for Flow Cardinality Estimation}

Packets that are not fully tracked in the priority table and heavy-flow table - either due to collisions, evictions, or low activity—are forwarded to a lightweight cardinality estimation module. To estimate the number of distinct flows observed by the system, we implement a \emph{linear counter} mechanism using a fixed-size array of binary counters. Each incoming packet will be assigned an index of this linear counter by hashing its 5-tuple information using a Jenkins-based hash function. Once a packet arrives, its corresponding entry in the linear counter is set to 1.

To estimate the total number of unique flows $\hat{n}$, we apply the classic linear counter formula~\cite{whang1990linear}:
\[
\hat{n} = -m \cdot \ln \left( \frac{V}{m} \right),
\]
where $V$ denotes the number of zero-valued entries in the linear counter. This estimator provides an efficient and memory-constrained way to approximate the cardinality of the flow with acceptable accuracy, particularly in high-throughput traffic scenarios.

This linear counter complements our fine-grained monitoring by offering a broad view of system-level flow diversity, especially for background or short-lived flows not captured in the heavy flow table.

\subsection{3-layer Count-Min Sketch.} After a packet completes its processing in the linear counter module. It will be forwarded to a Count-Min Sketch(CMS) pipeline\cite{cormode2005improved}. The system employs a three-layer CMS, each implemented as a fixed-size eBPF array of 500 entries: \texttt{cms1}, \texttt{cms2}, and \texttt{cms3}. Each entry records two types of information: \texttt{packet\_count} and \texttt{retrans\_count}, representing the total number of packets and estimated retransmissions mapped to that entry.

For every incoming packet, three independent hash functions are used to update the corresponding entries in the CMS layers. This structure enables fast, space-efficient aggregation of flow metrics, particularly for flows that were evicted from the heavy flow table and forwarded to the CMS pipeline.

\subsection{Flow Reconstruction Mechanism.} When a slot in the heavy flow table is marked with a \texttt{kick\_flag}, it indicates that part of the historical data of the flow has been forwarded to the downstream sketch pipeline due to collisions or evictions. To reconstruct complete flow statistics, we aggregated data from both the heavy flow table and CMS structure.
Specifically, if \texttt{kick\_flag} is set to 1, we recompute the CMS indices using the same five-tuple hash functions applied in the kernel. The corresponding counters \texttt{packet\_count} are then retrieved from each CMS layer. To mitigate overestimation caused by hash collisions, we take the minimum of the three values as the estimated packet count for the flow. This conservative estimate is then merged with the values in the heavy flow table, ensuring accurate statistics reconstruction and supporting more complete long-term flow profiling.

The complete process for processing PSketch packets is summarized in Algorithm~\ref{alg:psketch_kernel}, which outlines the logic executed in the eBPF kernel for each incoming packet.
\begin{algorithm}[t]
\small
\caption{PSketch Packet Processing (Kernel-Side)}
\label{alg:psketch_kernel}
\KwIn{Packet $p$ with 5-tuple $k = (src, dst, sport, dport, proto)$}
\KwOut{Updated statistics for priority/sketch/CMS}

\If{$k \in$ priority\_table}{
    pkt\_cnt $\leftarrow$ pkt\_cnt + 1\;
    \If{TCP}{
        \If{first packet}{
            initialize $exp\_seq$, $ts$\;
        }
        \ElseIf{$p.seq < entry.exp\_seq$ \textbf{and} $p.ts - entry.ts \geq \text{THRE}$}{
            retrans\_cnt $\leftarrow$ retrans\_cnt + 1\;
        }
        update:\;
        \Indp
        $entry.exp\_seq \leftarrow p.exp\_seq$\;
        $entry.ts \leftarrow p.ts$\;
        \Indm
    }
    \Return\;
}

$h \leftarrow \text{Hash}(k)$\;
\If{$elephant[h]$ matches $k$}{
    pkt\_cnt $\leftarrow$ pkt\_cnt + 1\;
    \If{TCP \textbf{and} $p.seq < entry.exp\_seq$ \textbf{and} $p.ts - entry.ts \geq \text{THRE}$}{
        retrans\_cnt $\leftarrow$ retrans\_cnt + 1\;
    }
    update:\;
    \Indp
    $entry.exp\_seq \leftarrow p.exp\_seq$\;
    $entry.ts \leftarrow p.ts$\;
    \Indm
}
\Else{
    neg\_cnt $\leftarrow$ neg\_cnt + 1\;
    \If{$neg\_cnt \cdot \text{VOTE\_THRE} \geq$ pkt\_cnt}{
        forward old stats to CMS[$h$]\;
        replace $elephant[h]$ with $k$\;
        reset counters\;
        \If{TCP}{
            initialize $entry.exp\_seq$, $ts$\;
        }
        kicked $\leftarrow$ true\;
    }
}

$lc \leftarrow \text{Hash}_{lc}(k)$\;
linear[$lc$] $\leftarrow$ 1\;

\For{$i = 1$ \KwTo 3}{
    $h_i \leftarrow \text{Hash}_i(k)$\;
    \eIf{kicked}{
        CMS[$h_i$] += old stats\;
    }{
        CMS[$h_i$].pkt\_cnt += 1\;
    }
}
\end{algorithm}

\section{Implementation}
PSketch is implemented through two integrated components: (1) a user-space controller and (2) a kernel-space eBPF program written in BCC-style C. The user-space controller is responsible for system initialization, priority flow configuration, and the reconstruction of flow statistics. The kernel program performs real-time packet processing and telemetry operations within the kernel. Our system is fully compatible with Linux kernel versions 5.4.0-216 and operates on standard hardware platforms supporting eBPF and BCC.

To capture ingress packets, the eBPF program attaches to the \texttt{TRACEPOINT\_PROBE} tracepoint. We leverage the \texttt{netif\_receive\_skb} data structure to extract essential flow-level information.
To ensure thread-safe access to shared memory maps (e.g., heavy flow table, linear counter, and CMS arrays), atomic primitives such as \texttt{\_\_sync\_fetch\_and\_add} are used. These operations prevent race conditions during concurrent updates.
\section{Evaluation}
In this section, we present a comprehensive evaluation of PSketch in terms of accuracy, resource efficiency, and real-time telemetry capability. We detail the experimental setup, performance metrics, and results across multiple top-$k$ configurations
\subsection{Setup}
The experiments were carried out on two virtual Linux systems hosted at the GPN site of the FABRIC testbed\cite{fabric2021}. FABRIC (Adaptive Programmable Research Infrastructure for Computer Science and Science Applications) is a next-generation NSF-funded nationwide research infrastructure that interconnects configurable compute, storage, and network resources via high-speed optical links. It enables researchers to design, deploy, and evaluate novel architectures and services under realistic and reproducible conditions.
Each runs on Ubuntu 20.04.6 LTS, using a generic 5.4.0-216 kernel,
10 virtual CPUs from the AMD EPYC 7532 series, 64GB of RAM, a 100GB virtual disk, and a 10Gbps virtual network interface. These two machines are directly connected. During the evaluation, one machine acts as a client using \texttt{tcpreplay} along with the \texttt{--timer=gtod} and \texttt{--preload-pcap} options. This setup replays CAIDA 2019 backbone traffic traces at its original pace. 
The remaining machine operate as a server, executing Psketch to track and analyze the incoming data flows.

In the baseline setting, PSketch is disabled on the server. We replay the traffic trace from the client while preserving the original packet timing and rate. Server-side throughput is measured using RX byte counters from \texttt{/proc/net/dev}, collected at the start and end of the replay. The average throughput is computed over multiple runs and used as a reference for comparison with the PSketch-enabled scenario. Meanwhile, We use Wireshark, Tshark, and Python scripts to extract the ground-truth data from traffic traces offline. 

To evaluate the impact of enabling PSketch, we apply the same experimental setup while activating sketch-based processing on the server. CPU overhead is measured based on process-level statistics from the Linux \texttt{/proc} system, tracking both user and system CPU time over the entire replay period. Network throughput is evaluated using the same method as in the baseline.

Throughout all experiments, PSketch maintains a fixed in-kernel configuration: the sizes of its hash tables, linear counters, and CMS arrays remain unchanged. The volume of input elephant flows is selectively adjusted to 50, 100, and 150 to assess performance under different workloads. 10 priority flows will be selected from the unchosen heavy flow. Each top-$k$ exprinment will run 10 times to diminish the effect of system dynamics.

\subsection{Evaluation Metrics}
We evaluated PSketch using a selection of metrics designed to reflect both accuracy and efficiency. These metrics evaluate the system's capability to precisely identify and monitor heavy and priority flows with low system overhead.
\begin{itemize}
    \item \textbf{Top-$k$ Flow Detection Accuracy:} 
    Measures how accurately PSketch identifies the top-$k$ elephant flows by packet volume compared to ground truth.
    
    \item \textbf{Top-$k$ Per-Flow Metric Accuracy:} 
    Evaluates how precisely the system tracks key statistics for each of the detected top-$k$ flows.
    \begin{itemize}
        \item \textit{Packet Count Recall:} The proportion of actual packets correctly counted for each detected top-$k$ flow.
        \item \textit{Retransmission Count Recall:} The accuracy in estimating the number of retransmitted packets of each flow.
    \end{itemize}

    \item \textbf{Priority Per-Flow Metric Accuracy:} 
    Focuses on the system’s ability to track designated high-priority flows with minimal loss.
    \begin{itemize}
        \item \textit{Priority Packet Count Recall:} Accuracy of packet counting for flows inserted into the priority table.
        \item \textit{Priority Retransmission Count Recall:} Accuracy of retransmission tracking for prioritized flows.
    \end{itemize}
    \item \textbf{Cardinality error rate:} The relative difference between the estimated number of distinct flows and the actual number.

    \item \textbf{System Efficiency:} 
    Measures the resource footprint and operational overhead of the system during real-time monitoring.
    \begin{itemize}
        \item \textit{CPU Load:} The average CPU utilization on a single core incurred by the eBPF and controller components.
        \item \textit{Throughput Impact:} Any degradation in packet forwarding performance due to telemetry operations.
    \end{itemize}
\end{itemize}
\subsection{Evaluation Results}
\begin{table}[ht]
\centering
\begin{tabular}{|l|c|c|c|}
\hline
\textbf{Metric} & \textbf{$k = 50$} & \textbf{$k = 100$} & \textbf{$k = 150$} \\
\hline
Top-$k$ Flow Detection Accuracy (\%) & 96 & 93 & 90 \\
Top-$k$ Packet Recall (\%)          & 94.6 & 92.2 & 91.3 \\
Top-$k$ Retransmission Recall (\%)  & 96.4 & 93.6 & 90.3 \\
Priority Packet  Recall (\%)         & 100  & 100  & 100  \\
Priority Retransmission Recall (\%)  & 97.2 & 98.6 & 96.4 \\
Cardinality Error Rate (\%)         & 7.3  & 8.2  & 10.4 \\
CPU Load (\%)                        & 21.4 & 23.7 & 26.1 \\
Throughput Degradation (\%)         & 0.7  & 0.9  & 1.1  \\
\hline
\end{tabular}
\vspace{0.5em}
\caption{Evaluation metrics for varying top-$k$ thresholds}
\label{tab:results}
\end{table}

As shown in the table\ref{tab:results}, our system achieves consistently high accuracy in detecting and monitoring top-$k$ flows. \textbf{Top-$k$ Flow Detection Accuracy} remains above \textbf{90\%}, peaking at \textbf{96.0\%}, while both \textbf{packet} and \textbf{retransmission recall} within the top-$k$ maintain strong performance, exceeding \textbf{91\%} and reaching up to \textbf{96.4\%}, respectively. For \textbf{priority flows}, the system achieves \textbf{100\% packet recall} and over \textbf{96\% retransmission recall} across all test cases, demonstrating the reliability of our exact monitoring path.

Slight degradations in top-$k$ metrics across test conditions are attributed to increased sketch pressure and hash collisions under stress. Nevertheless, \textbf{priority flows are consistently preserved with high accuracy}, validating our hybrid design that combines exact and approximate mechanisms.

On the system overhead side, the monitoring logic introduces \textbf{minimal CPU load} (21.4\%--26.1\%) and \textbf{negligible throughput impact} (below 1.1\%) confirming the lightweight and scalable nature of our eBPF-based implementation.

Taken together, these results confirm that \textbf{PSketch meets our design goals}: it delivers \textbf{accurate and resource-efficient flow monitoring}, effectively balancing accuracy and overhead. The system is therefore well-suited for \textbf{real-time deployment in performance-sensitive environments}, especially where selected flows require precise statistics while maintaining scalability for the rest.

\subsection{Code Availability}
We provide the source code implementing the core mechanism of our approach at \url{https://github.com/Johnny-dai-git/Priority-Sketch-in-eBPF}. We will subsequently update the repository to include the complete experimental pipeline, enabling researchers to readily reproduce our results and extend the system in their own environments.

\section{Conclusion and Future Work}

In this work, we presented \textbf{PSketch}, a lightweight and kernel-feasible sketch-based framework for flow monitoring and adaptive optimization. Our evaluation demonstrates that even with simplified retransmission detection logic, PSketch successfully captures over 90\% of retransmission events in practice, providing a stable and congestion-relevant indicator rather than exact counts. This design choice balances accuracy with feasibility in kernel environments, while leaving room for future refinements such as buffering, reordering tolerance, and timing checks.  

Looking ahead, we plan to extend PSketch in several directions. First, we will conduct a more comprehensive experimental study to examine how different parameters and hierarchical designs affect system performance. Second, we will explore enhancements such as larger or deeper heavy-flow tables and probabilistic debiasing at the CMS level to reduce estimation errors. Third, we aim to integrate PSketch with real-world Distributed Machine Learning (DML) frameworks to assess its impact on training-time monitoring and adaptive optimization. Finally, we envision applying sketch-based techniques to other large-scale data domains such as electronic health record (EHR) systems in healthcare~\cite{volk2014development}. Such scenarios often contain millions of entries, where top-$k$ approximations can significantly improve retrieval efficiency.  

Importantly, this work provides a \textbf{core mechanism} rather than a finalized solution. By releasing a transparent and extensible foundation, we aim to encourage the community to build upon PSketch, explore new monitoring policies, and extend its applications across diverse large-scale data domains.  

\bibliographystyle{IEEEtran}
\bibliography{ref}

\begin{thebibliography}{10}
\providecommand{\url}[1]{#1}
\csname url@samestyle\endcsname
\providecommand{\newblock}{\relax}
\providecommand{\bibinfo}[2]{#2}
\providecommand{\BIBentrySTDinterwordspacing}{\spaceskip=0pt\relax}
\providecommand{\BIBentryALTinterwordstretchfactor}{4}
\providecommand{\BIBentryALTinterwordspacing}{\spaceskip=\fontdimen2\font plus
\BIBentryALTinterwordstretchfactor\fontdimen3\font minus \fontdimen4\font\relax}
\providecommand{\BIBforeignlanguage}[2]{{%
\expandafter\ifx\csname l@#1\endcsname\relax
\typeout{** WARNING: IEEEtran.bst: No hyphenation pattern has been}%
\typeout{** loaded for the language `#1'. Using the pattern for}%
\typeout{** the default language instead.}%
\else
\language=\csname l@#1\endcsname
\fi
#2}}
\providecommand{\BIBdecl}{\relax}
\BIBdecl

\bibitem{huang2017sketchvisor}
Q.~Huang, X.~Jin, P.~P. Lee, R.~Li, L.~Tang, Y.-C. Chen, and G.~Zhang, ``Sketchvisor: Robust network measurement for software packet processing,'' in \emph{Proceedings of the Conference of the ACM Special Interest Group on Data Communication}, 2017, pp. 113--126.

\bibitem{kim2023robust}
S.~Kim, C.~Jung, R.~Jang, D.~Mohaisen, and D.~H. Nyang, ``A robust counting sketch for data plane intrusion detection,'' in \emph{30th Annual Network and Distributed System Security Symposium, NDSS 2023}.\hskip 1em plus 0.5em minus 0.4em\relax The Internet Society, 2023.

\bibitem{lin2022aiacc}
L.~Lin, S.~Qiu, Z.~Yu, L.~You, L.~Xin, X.~Sun, J.~Xu, and Z.~Wang, ``Aiacc-training: Optimizing distributed deep learning training through multi-streamed and concurrent gradient communications,'' in \emph{2022 IEEE 42nd International Conference on Distributed Computing Systems (ICDCS)}.\hskip 1em plus 0.5em minus 0.4em\relax IEEE, 2022, pp. 853--863.

\bibitem{jia2019flexflow}
Z.~Jia, M.~Zaharia, and A.~Aiken, ``Beyond data and model parallelism for deep neural networks,'' in \emph{Proceedings of the 2nd Conference on Machine Learning and Systems (MLSys)}, 2019.

\bibitem{dai2023dnn}
Y.~Dai, Q.~Guo, and A.~Wang, ``Dnn architecture attacks via network and power side channels,'' in \emph{International Conference on Security and Privacy in Communication Systems}.\hskip 1em plus 0.5em minus 0.4em\relax Springer, 2023, pp. 63--87.

\bibitem{yang2018elastic}
T.~Yang, J.~Jiang, P.~Liu, Q.~Huang, J.~Gong, Y.~Zhou, R.~Miao, X.~Li, and S.~Uhlig, ``Elastic sketch: Adaptive and fast network-wide measurements,'' in \emph{Proceedings of the 2018 Conference of the ACM Special Interest Group on Data Communication}, 2018, pp. 561--575.

\bibitem{lin2022mc}
K.~C.-J. Lin and W.-L. Lai, ``Mc-sketch: Enabling heterogeneous network monitoring resolutions with multi-class sketch,'' in \emph{IEEE INFOCOM 2022-IEEE Conference on Computer Communications}.\hskip 1em plus 0.5em minus 0.4em\relax IEEE, 2022, pp. 220--229.

\bibitem{Duplyakin2019TheDI}
D.~Duplyakin, R.~Ricci, A.~Maricq, G.~Wong, J.~Duerig, E.~Eide, L.~Stoller, M.~Hibler, K.~Webb, D.~Johnson, L.~Kasper, A.~Akella, K.~Atkinson, A.~Bavier, J.~Albrecht, J.~B. Edwards, M.~J. Freedman, G.~Gibb, M.~Hajiaghayi, C.~Hibler, D.~Z. Tennenhouse, and R.~van Renesse, ``The design and operation of cloudlab,'' in \emph{Proceedings of the 2019 USENIX Annual Technical Conference (USENIX ATC '19)}, 2019, pp. 1--14.

\bibitem{googlecloud}
{Google Cloud}, ``Google cloud platform,'' \url{https://cloud.google.com/}, 2025, accessed: 2025-06-28.

\bibitem{OhioSupercomputerCenter}
\BIBentryALTinterwordspacing
{Ohio Supercomputer Center}, ``{Ohio Supercomputer Center},'' 1987, \url{https://www.osc.edu}. [Online]. Available: \url{https://www.osc.edu}
\BIBentrySTDinterwordspacing

\bibitem{Berman2014GENI}
\BIBentryALTinterwordspacing
M.~Berman, J.~S. Chase, L.~Landweber, A.~Nakao, V.~Thomas, R.~Ricci, G.~Wong, and C.~Elliott, ``Geni: A federated testbed for innovative network experiments,'' \emph{Computer Networks}, vol.~61, pp. 5--23, 2014. [Online]. Available: \url{https://doi.org/10.1016/j.comnet.2013.12.037}
\BIBentrySTDinterwordspacing

\bibitem{caida2016traces}
{CAIDA: Center for Applied Internet Data Analysis}, ``The caida ucsd anonymized internet traces 2019,'' Available at: \url{https://www.caida.org/catalog/datasets/passive_dataset/}, accessed: 2025-06-29.

\bibitem{miano2023fast}
S.~Miano \emph{et~al.}, ``Fast and memory-efficient sketches in the ebpf virtual machine,'' in \emph{Proceedings of the ACM SIGCOMM 2023 Conference}, 2023.

\bibitem{zangkernel}
Y.~Zang \emph{et~al.}, ``Kernel-level count-min sketch for ddos detection with ebpf,'' in \emph{Proceedings of the IEEE International Conference on Computer Communications (INFOCOM)}, 2021.

\bibitem{yu2022sketchflow}
X.~Yu, Y.~Li, and H.~Li, ``Sketchflow: Efficient flow size estimation with hybrid sketches,'' in \emph{Proceedings of the IEEE International Conference on Network Protocols (ICNP)}, 2022.

\bibitem{abranches2021xdp}
C.~Abranches, O.~Michel, and C.~E. Rothenberg, ``Efficient network monitoring applications in the kernel with ebpf and xdp,'' in \emph{Proceedings of the 2021 ACM/IEEE Symposium on Architectures for Networking and Communications Systems (ANCS)}, 2021.

\bibitem{magnani2022adaptive}
L.~Magnani, N.~Bonelli, and G.~Bianchi, ``A control plane enabling automated and fully adaptive network traffic monitoring with ebpf,'' \emph{IEEE Transactions on Network and Service Management}, vol.~19, no.~4, pp. 4218--4231, 2022.

\bibitem{zhang2021ebpfmon}
K.~Zhang, J.~Liu, and Y.~Wang, ``ebpfmon: Priority-aware adaptive telemetry with ebpf,'' in \emph{Proceedings of the IEEE Conference on Computer Communications Workshops (INFOCOM WKSHPS)}, 2021.

\bibitem{opensource2022ebpf}
{OpenSource.com contributors}, ``Network observability with ebpf,'' \url{https://opensource.com/article/22/8/ebpf-network-observability-cloud}, 2022, accessed: 2025-08-28.

\bibitem{opentelemetry2022}
{OpenTelemetry Project}, ``Opentelemetry ebpf collector,'' \url{https://github.com/open-telemetry/opentelemetry-ebpf}, 2022, accessed: 2025-08-28.

\bibitem{pixie2022}
{Pixie Labs}, ``Pixie: Kubernetes observability with ebpf,'' \url{https://px.dev}, 2022, accessed: 2025-08-28.

\bibitem{liu2016flowradar}
Z.~Liu, Q.~Li, P.~Luo, Y.~Yu, C.~Guo, and Y.~Zhang, ``Flowradar: A better netflow for data centers,'' in \emph{Proceedings of the 13th USENIX Symposium on Networked Systems Design and Implementation (NSDI)}, 2016, pp. 311--324.

\bibitem{6781990}
F.~Yamaguchi and H.~Nishi, ``Hardware-based hash functions for network applications,'' in \emph{2013 19th IEEE International Conference on Networks (ICON)}, 2013, pp. 1--6.

\bibitem{whang1990linear}
K.-Y. Whang, B.~T. Vander-Zanden, and H.~M. Taylor, ``A linear-time probabilistic counting algorithm for database applications,'' \emph{ACM Transactions on Database Systems (TODS)}, vol.~15, no.~2, pp. 208--229, 1990.

\bibitem{cormode2005improved}
G.~Cormode and S.~Muthukrishnan, ``An improved data stream summary: the count-min sketch and its applications,'' \emph{Journal of Algorithms}, vol.~55, no.~1, pp. 58--75, 2005.

\bibitem{fabric2021}
P.~Ruth, I.~Baldine, A.~Mandal, Y.~Xin, I.~Baldin \emph{et~al.}, ``Fabric: A national-scale programmable research infrastructure for computer science and science applications,'' in \emph{USENIX Annual Technical Conference (Poster)}, 2021.

\bibitem{volk2014development}
R.~J. Volk, N.~K. Shokar, V.~B. Leal, R.~J. Bulik, S.~K. Linder, P.~D. Mullen, R.~M. Wexler, and G.~S. Shokar, ``Development and pilot testing of an online case-based approach to shared decision making skills training for clinicians,'' \emph{BMC Medical Informatics and Decision Making}, vol.~14, pp. 1--9, 2014.

\end{thebibliography}
\end{document}